\documentclass[a4paper,11pt]{article}
\usepackage{jinstpub} 
\usepackage{lineno}
\usepackage[sicmds,freestanding]{hepunits}
\usepackage{todonotes}
\usepackage{xcolor}
\usepackage{booktabs}
\usepackage{subcaption} 

\usepackage[numbers,sort&compress]{natbib}

\title{\boldmath AI-Enhanced Self-Triggering for Extensive Air Showers: Performance and FPGA Feasibility}

\author{Qader Dorosti}
\affiliation{Center for Particle Physics Siegen, \\ 
Department für Physik, Universität Siegen,\\
Walter-Flex-Str. 3, 57072 Siegen, Germany}

\emailAdd{dorosti@hep.physik.uni-siegen.de}

\abstract{
Autonomous self-triggering for radio detection of extensive air showers remains a long-standing challenge, particularly in environments dominated by strong and variable radio-frequency interference. Current radio arrays usually rely on external particle-detector triggers: while this lowers thresholds for vertical showers, it excludes very inclined events, where radio detection is uniquely powerful because the particle cascade is absorbed while the radio pulse remains measurable. In this work, I present a proof-of-principle study showing that deep-learning-based triggering can overcome these limitations, operating robustly under realistic high-interference conditions and within the strict latency constraints of large-scale observatories. Using measured noise traces combined with simulated cosmic-ray pulses, a fully convolutional network was trained and optimised for MHz-scale trigger-trial rates, and its performance was compared to a simple threshold-based trigger, which proved markedly less efficient at the same false-positive rate. The trained model was then quantised with \texttt{HLS4ML} and synthesised with \texttt{Vitis HLS} for multiple FPGA targets. The quantised implementation preserves the floating-point model’s physics performance (\(\mathrm{AUC}_{\mathrm{float}}=0.997\), \(\mathrm{AUC}_{\mathrm{quant}}=0.996\)) while achieving microsecond-scale inference latencies. These results show that modern FPGA-based AI inference can provide low-latency, radio-only triggering for next-generation cosmic-ray observatories, particularly in regions with strong and variable radio backgrounds. This capability unlocks access to weak and highly inclined air-shower signals---even in noisy environments---thereby broadening the energy range, extending sky coverage, and opening the path toward ultra-high-energy neutrino detection with ground-based radio arrays.
}


\begin{document}
\maketitle
\flushbottom

\section{Introduction}
\label{sec:intro}
Ultra-high-energy cosmic rays (UHECRs) are primarily detected using large-scale ground-based observatories that employ a combination of particle detectors, fluorescence telescopes, and radio antennas~\cite{SCHRODER20171}. Each technique provides complementary insights into extensive air showers (EAS), enabling detailed studies of cosmic-ray composition, energy spectrum, and arrival direction. Among these methods, radio detection has gained increasing attention due to its cost-effectiveness, full-sky coverage, and near-continuous operation~\cite{HUEGE20161}. By measuring the radio emission generated primarily by geomagnetic deflection~\cite{geomEffect} and the charge-excess effect~\cite{AskEffect} in the EAS, radio arrays can reconstruct key shower parameters, such as the atmospheric depth of the shower maximum ($X_{\text{max}}$) and primary energy, with high precision~\cite{Abdul_Halim2024-ll,Aab2016-fq}.

A distinct advantage of radio detection is its sensitivity to inclined air showers, where traditional particle detector arrays struggle due to the attenuation of the electromagnetic shower component in the atmosphere~\cite{Aab2016-fq}. However, despite these benefits, current radio detection techniques rely on external triggers from other detector systems, such as surface or fluorescence detectors, limiting their autonomy and increasing operational complexity. The development of an effective self-triggering system is therefore crucial to fully exploit the potential of radio detection for cosmic-ray observations~\cite{Schmidt2011-xy}.

Previous attempts to implement self-triggering in radio arrays, such as those by CODALEMA and AERA, relied on conventional techniques like narrowband radio frequency interference (RFI) suppression and threshold-based triggering~\cite{Torres_Machado2013-my,KELLEY2013133}. However, these methods proved ineffective in distinguishing cosmic-ray-induced radio pulses from transient noise sources, leading to a high rate of false positives. As a result, these efforts faced significant challenges and did not produce established scientific results, underscoring the need for more advanced signal-processing strategies. 


Notably, the ANITA~\cite{Hoover2010-pa} and TREND~\cite{Charrier2019-ti,Ardouin2011-hu} experiments successfully implemented autonomous radio self-triggering. 
ANITA, a balloon-borne radio interferometer, operated in the stratosphere, where the continuous background was comparatively quiet but impulsive anthropogenic RFI events were still present and were identified and rejected using polarisation information in subsequent analysis. TREND, by contrast, deployed a ground-based array in a remote, radio-quiet region of China, where ambient noise levels were exceptionally low. In both cases, comparatively favourable noise conditions were crucial in enabling effective self-triggering with simple threshold-based approaches—conditions that are significantly more challenging in the high-interference environments considered in this work.

Recent developments in modelling EAS radio emissions, combined with experimental and simulation studies, have provided a deeper understanding of characteristic pulse shapes and ground radiation patterns~\cite{HUEGE20161}. This progress has paved the way for more sophisticated signal-processing techniques, including artificial intelligence (AI)-driven approaches for enhancing radio-based self-triggering. Recent studies show that deep learning models, including convolutional neural networks (CNNs) and recurrent neural networks (RNNs), can effectively distinguish cosmic-ray signals from background noise, significantly reducing detection thresholds. These techniques have been successfully applied to identify air-shower radio pulses in offline analyses~\cite{Schroder2024-ao,BEZYAZEEKOV201589,Erdmann2019-az}, demonstrating their potential to enhance signal selection. Recent efforts, such as those in the GRAND experiment, continue to investigate self-triggering using AI-based methods, including CNNs. While preliminary results indicate that these methods can reject at least 40\% of background noise at a 90\% signal selection efficiency~\cite{Correa2024-ap}, achieving real-time deployment remains challenging. Self-triggering requires ultra-low latency processing to ensure the identification of cosmic-ray signals within microseconds of their arrival, while also maintaining a false-positive trigger rate at the sub-\SI{}{\Hz} level, a constraint that traditional software-based neural networks struggle to meet.

To address the challenge of self-triggering in cosmic-ray radio signal detection, AI-accelerated signal processing implemented on Field-Programmable Gate Arrays (FPGAs) offers a viable solution. FPGAs provide exceptional parallel processing capabilities and low-latency inference, making them ideal for real-time feature extraction and classification of radio pulses. Similar approaches have been explored in high-energy physics experiments, such as the ATLAS Tile Calorimeter, where deep learning algorithms implemented on FPGAs have demonstrated significant improvements in real-time signal reconstruction under high pileup conditions~\cite{Arciniega_2019,Chiedde_2022}. These advancements highlight the potential of FPGA-based AI techniques in handling complex, high-rate signal environments, which are also crucial for enhancing the efficiency and accuracy of cosmic-ray self-triggering.

In this work, I investigate the feasibility of FPGA‑based AI self‑triggering for radio detection of EAS.  
I begin by training and evaluating a fully convolutional network on GPU using a dataset that combines \textit{experimentally measured background noise} with \textit{simulated cosmic‑ray pulses} generated by the MGMR3D framework~\cite{Scholten2008-fk,Werner2008-ad,Scholten2019-va,Scholten2023-hc}.  
The trained model is then quantised with \texttt{hls4ml}~\cite{fastml_hls4ml,Duarte:2018ite,Aarrestad:2021zos} to fixed‑point precision and synthesised using \texttt{Vitis HLS~2023.2}~\cite{vitis_ai_2025} for multiple FPGA targets, allowing direct assessment of post‑synthesis latency and resource utilisation.  

By leveraging measured noise from a high‑interference environment, the approach is designed for robustness in realistic deployment conditions.  
I address key considerations for FPGA‑based AI triggers, including quantisation, latency optimisation, and DSP‑constrained architectures, and demonstrate that sub‑microsecond inference latencies are achievable within realistic resource budgets.  

The implementation of AI‑accelerated FPGA‑based self‑triggering represents a significant advancement in radio detection technology.  
By reducing reliance on external detector triggers and enabling high‑efficiency, low‑latency signal identification, this work opens the path toward autonomous operation of next‑generation radio observatories.  
It marks the first demonstration, in the context of radio‑detector self‑triggering, of a complete workflow from model training to FPGA‑synthesised deployment using \texttt{hls4ml}, providing a solid foundation for future developments in real‑time cosmic‑ray radio detection.

\section{Experimental Setup}  
The dataset used in this study consists of experimentally recorded noise traces collected in the physics campus of the University of Siegen. The measurements were performed using a butterfly antenna connected to a low-noise amplifier with 30 dB gain, which was further linked to a digital oscilloscope. These recordings capture real-world transient noise sources, including atmospheric and anthropogenic contributions. 

To focus on the frequency range relevant for air shower radio detection, the recorded signals were digitally bandpass-filtered between 30 and 80 MHz. Furthermore, data was sampled at \SI{250}{\MHz} to ensure comparability with real-world cosmic-ray radio detectors, such as AERA and Auger RD, which operate at \SI{180}{\MHz} and \SI{250}{\MHz}, respectively. This adjustment reflects the practical constraints of large-scale detectors, where lower sampling rates are used due to budget and hardware limitations, ensuring the applicability of the method to future radio arrays.

The dataset used in this study was collected in an urban environment, where background noise levels are higher and more variable compared to remote cosmic-ray observatories. To capture a diverse range of real-world noise conditions, data was collected over approximately two hours, yielding 200 million samples. Data was continuously streamed, with traces read out at a rate of one per second. Given the high variability of the urban noise environment, the dataset presents a challenging test case for AI-driven signal detection methods.

\begin{figure}[htbp]
\centering
\includegraphics[width=0.95\textwidth]{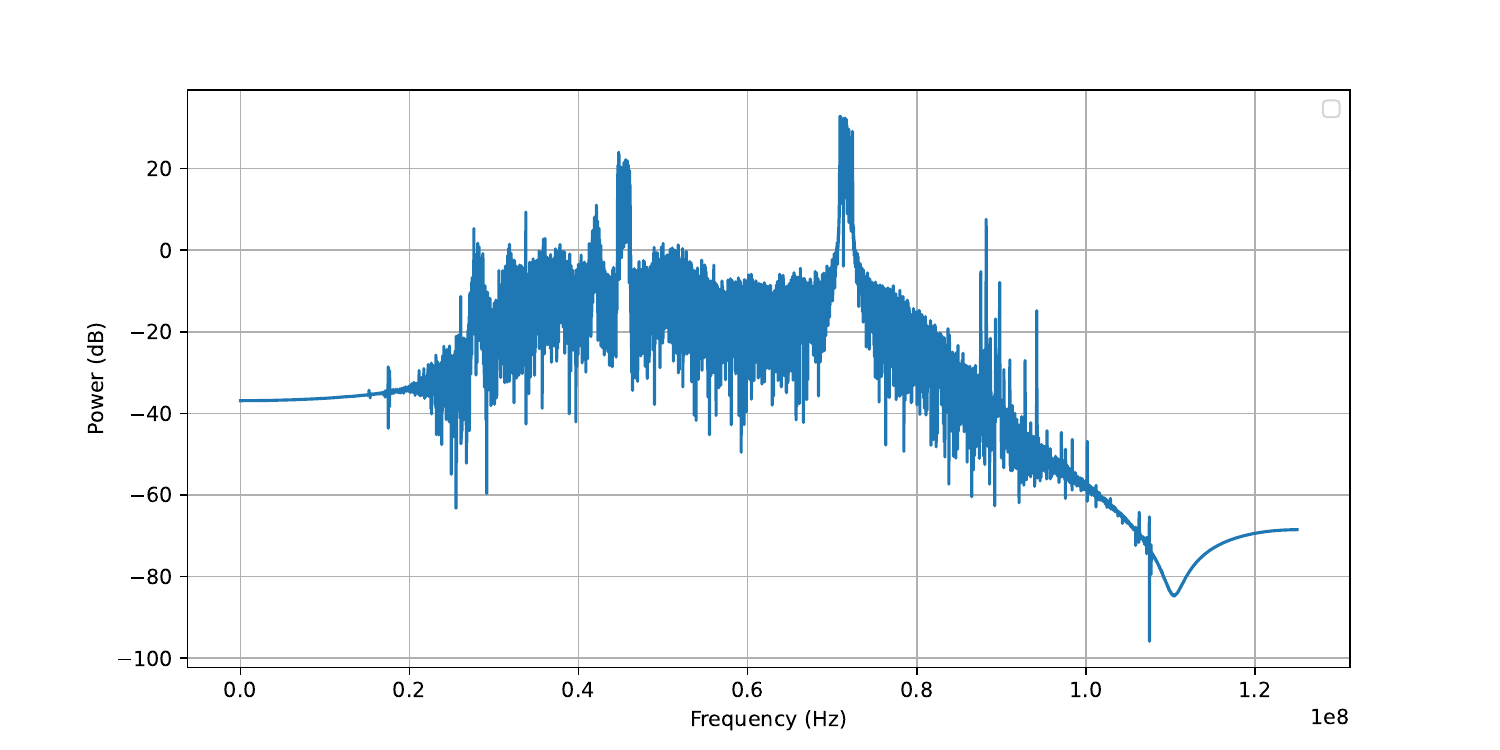}
\caption{Power spectrum of the recorded signal after applying a bandpass filter in the range of 30–\SI{80}{\MHz}. Two prominent narrowband RFI sources are observed at approximately 45 MHz and 71 MHz.}
\label{fig:spectrum}
\end{figure}

This setting introduces additional challenges, including increased transient interference and strong RFI. Two prominent narrowband RFI sources were identified at approximately \SI{45}{\MHz} and \SI{71}{\MHz}, along with other narrowband and transient RFI. The power spectrum of the recorded traces, shown in Figure~\ref{fig:spectrum}, was obtained after applying a digital bandpass filter in the range of 30 to \SI{80}{\MHz}. For the initial analysis, these narrowband RFIs were not suppressed to assess the model’s robustness in the presence of strong interference. The results presented in the following section reflect model performance when both training and testing were conducted on data sets containing these RFIs. 

To evaluate the AI model’s performance under realistic and diverse signal conditions, I generated air-shower radio pulses using the MGMR3D simulation framework~\cite{Scholten2008-fk,Werner2008-ad,Scholten2019-va,Scholten2023-hc}, which reproduces the time-domain waveform and frequency content expected from extensive air showers for a wide range of shower geometries and primary energies.
The simulated dataset spans zenith angles from $0^\circ$ to $85^\circ$ in $5^\circ$ steps, uniformly distributed random azimuths, and primary energies between \SI{10}{\exa\electronvolt} and \SI{200}{\exa\electronvolt}.
For each geometry, $X_{\mathrm{max}}$ was varied randomly between 600 and $\SI{800}{g\per\centi\meter\squared}$ to cover the natural spread observed in air-shower development.
Signals were simulated for antenna positions at distances of \SI{50}{\meter} to \SI{400}{\meter} from the shower core, ensuring realistic variation in pulse amplitude, shape, polarity, and spectral content across the dataset. In total, this resulted in the generation of over 78,000 simulated radio pulses.

These pulses, already encompassing a wide range of shapes and amplitudes from the simulated variation in energy, geometry, $X_{max}$ values and antenna–core distance, were injected at random times into measured noise traces from a high‑interference environment. To control the signal‑to‑noise ratio (SNR) within a user‑defined range—without performing an absolute amplitude conversion from simulated electric‑field strength to the measured detector response—I rescaled each simulated pulse relative to the noise power, calculated from a pre‑filtered version of the corresponding trace.
To preserve the statistical properties of the noise and avoid any distortion from repeated filtering, pulses were injected into the original (unfiltered) traces.
The same digital bandpass filter was then applied to both noise-only and signal-injected traces, ensuring identical processing and conditioning across the dataset.

By combining realistic MGMR3D-generated air-shower pulses with measured noise, this approach enables a robust and representative evaluation of AI-based triggering performance. It captures the natural diversity of pulse shapes expected in real events while providing controlled SNR conditions. This ensures that the results are directly relevant for estimating the achievable false‑positive rate under realistic and challenging RFI conditions.

\subsection{Training and Validation Datasets}
\label{sec:dataset_definition}
To train and evaluate the model, multiple datasets were used, each designed to fulfil a specific role in ensuring robust performance. The datasets were categorized into training, cross-validation, and independent validation sets to assess the model’s generalization and sensitivity. All noise traces are unique and are derived from experimental measurements, randomly selected from the full duration of the noise measurement campaign, whereas the signal-containing traces consist of injected pulses, as previously described. For the initial analysis, only a bandpass filter was applied, with no additional filtering. 

\begin{itemize}  
    \item \textbf{Training and Cross-Validation Datasets:}  
    The dataset used for training and model tuning was randomly split into two equal parts:  
    \begin{itemize}  
        \item \textbf{Training Dataset (50\%):} Comprised \textbf{15,000 time series traces}, each containing 128 samples recorded at a sampling rate of \SI{250}{\MHz}, consistent with standard practices in radio-based air shower detection. It included an equal distribution of \textbf{5,000 pure noise traces} (\texttt{background}), \textbf{5,000 traces of simulated EAS pulses} (\texttt{pure signal}), and \textbf{5,000 traces with simulated EAS pulses injected into noise} (\texttt{signal}). This setup ensured that the model effectively learned to distinguish genuine signals from background noise.  
        
        \item \textbf{Cross-Validation Dataset (50\%):} Consisted of another \textbf{15,000 traces}, also evenly divided into \textbf{5,000 background traces}, \textbf{5,000 pure signal traces}, and \textbf{5,000 signal traces}. This dataset was used to fine-tune the model and monitor its generalization performance during training.  
    \end{itemize}  
    These datasets were randomly derived from a larger pool of available traces, ensuring an unbiased and representative sample distribution.  
      
    \item \textbf{Independent Validation Datasets:} Three independent validation datasets were used to assess the robustness of the model under different conditions.  
    \begin{itemize}  
        \item \textbf{Validation Dataset 1:} Contained \textbf{20,000 traces}, equally split into \textbf{10,000 background} and \textbf{10,000 signal} traces. This dataset was applied to evaluate the performance of the model on unseen data with an SNR distribution similar to the training set.  
        \item \textbf{Validation Dataset 2:} Consisted of \textbf{20,000 traces}, equally divided into \textbf{10,000 background traces} and \textbf{10,000 signal traces with extremely low-amplitude pulses}. This dataset was specifically designed to test the sensitivity of the model in low-SNR scenarios.  
        \item \textbf{Validation Dataset 3:} Comprised \textbf{1 million pure noise traces} to assist in evaluating the \textbf{false positive triggering rate}. This dataset provides a large statistical sample to quantify how frequently the model incorrectly classifies noise as a signal event.  
    \end{itemize}  
    
\end{itemize}  

These independent validation data sets were crucial to assessing the model's ability to generalize to previously unseen data and to reliably detect signals under varying noise conditions, particularly at low SNR.  

To quantify the SNR in a way that is robust to real detector conditions, I adopt a trace-wise definition using pure signal and pure noise traces (background) as inputs. Specifically, I define SNR as the squared peak amplitude of the Hilbert envelope of a pure signal trace, divided by the variance of a matched pure noise trace. This avoids ambiguity over whether the "signal" contains background or not. Formally, I compute
\begin{equation}
\label{eq:snr}
\mathrm{SNR} = \frac{A_\mathrm{peak}^2}{\sigma_\mathrm{noise}^2},
\end{equation}

where $A_\mathrm{peak}$ is the maximum amplitude of the analytic signal (obtained via the Hilbert transform) and $\sigma_\mathrm{noise}^2$ is the noise power computed as the variance of a pure noise trace. This approach emphasises the prominence of the signal over the background without being sensitive to specific waveform shapes.

To provide a baseline for background-only traces, I apply the same method to pure noise traces, treating them as if they contained a signal. Specifically, I compute the Hilbert envelope of each noise trace and take the maximum value as a hypothetical signal peak, then divide by the noise variance. This procedure captures spurious fluctuations in the noise that might mimic signal-like features. As a result, I obtain a distribution of apparent SNR values for background, which are typically lower than true signals but may overlap in some cases. This overlap justifies the use of logarithmic scaling and motivates the application of probabilistic, rather than deterministic, classification thresholds.

\begin{figure}[htbp]
\centering
\includegraphics[width=0.495\textwidth]{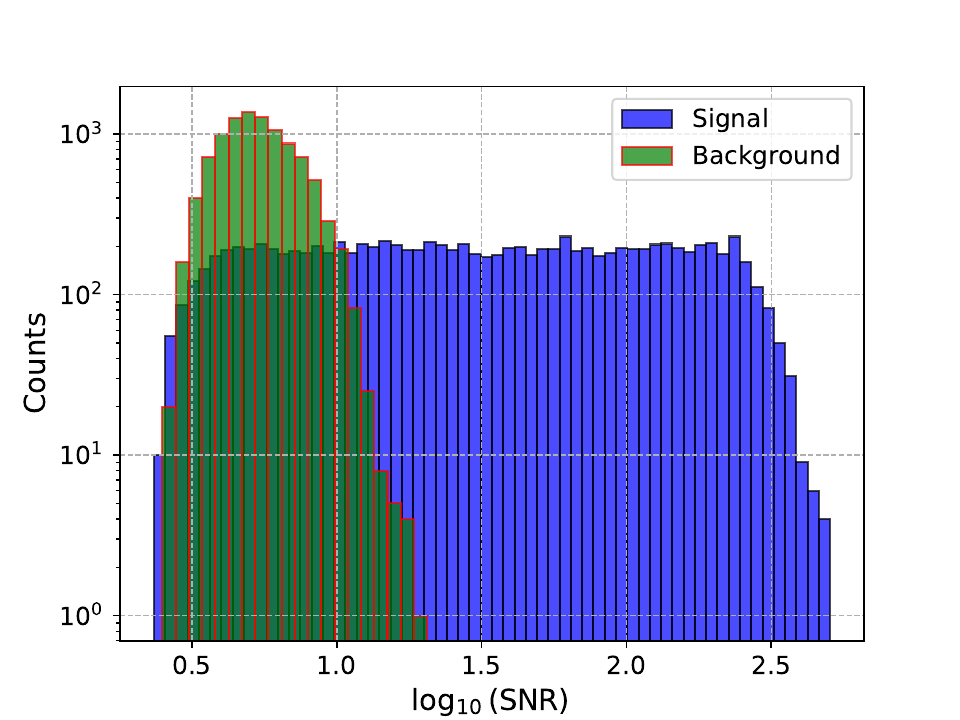}
\includegraphics[width=0.495\textwidth]{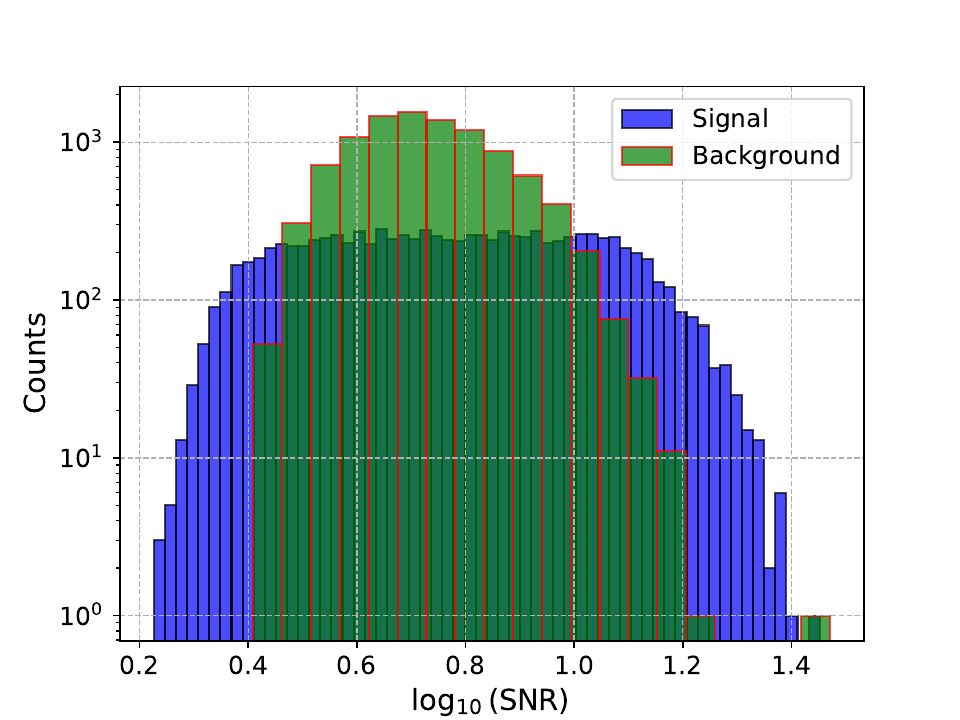}
\caption{SNR distributions for background and signal traces in Validation Dataset 1 (left) and Validation Dataset 2 (right). Validation Dataset 1 follows the distribution of the training dataset, while Validation Dataset 2 contains low-amplitude signals.}
\label{fig:snrDistr}
\end{figure}

Figure~\ref{fig:snrDistr} illustrates the SNR distributions for background and signal traces across Validation Dataset 1, which closely follows the distribution of the training dataset, and Validation Dataset 2, which consists of low-amplitude signals. This comparison provides insights into the differences in signal detectability across varying SNR conditions. In the SNR histograms, I report the base-10 logarithm of the computed SNR.

\section{Methodology} 

The goal of my approach is to design, train, and evaluate a neural network classifier capable of robustly detecting air‑shower radio pulses in realistic, high‑interference noise conditions, while ensuring compatibility with efficient FPGA deployment.
The methodology comprises two main components:
(1) the design of a fully convolutional network architecture optimised for both classification performance and hardware constraints, and
(2) the training and optimisation procedure used to achieve high detection efficiency with minimal false‑positive rate.

\subsection{Model Architecture}

The classification network is a fully convolutional 1-D architecture for radio traces with one input channel.  
It contains approximately \(1.5\times 10^4\) trainable parameters and was designed to maintain classification performance comparable to earlier floating-point models while ensuring that each convolutional or linear layer contains fewer than \num{4096} weights---a practical constraint arising from \texttt{Vitis HLS} synthesis failures when unrolling larger layers due to internal 12-bit address limitations in the memory architecture of \texttt{Vitis HLS}, which restrict access to weight arrays exceeding 4096 entries.

This constraint facilitates efficient FPGA implementation in later stages of the work without excessive on-chip memory use or pipeline latency.

The network consists of three convolutional blocks followed by global average pooling and a fully connected output layer.  
Batch normalisation (BN) is applied after each convolution, ReLU activations follow all but the final convolution in each block, and dropout is applied after each pooling stage.

The \textbf{first block} applies a 1D convolution with kernel size~3, 
1 input channel, and 16 output channels (\(\mathrm{stride}=1, \mathrm{padding}=1\)),
\begin{equation}
\mathbf{y} = \sigma\left(\mathrm{BN}\left(W^{(1)} * \mathbf{x}\right)\right), 
\quad W^{(1)} \in \mathbb{R}^{16 \times 1 \times 3},
\end{equation}
where \(\sigma\) denotes the ReLU activation. 
This is followed by max-pooling with factor~2 and dropout with rate~0.3.

The \textbf{second block} begins with a 1D convolution of kernel size~3 
that increases the number of channels from 16 to 32, followed by batch normalisation and ReLU activation:
\[
\mathbf{z} = \sigma\left(\mathrm{BN}\left(W^{(2)} * \mathbf{x}\right)\right), \quad W^{(2)} \in \mathbb{R}^{32 \times 16 \times 3}.
\]
This is followed by a residual sub-network composed of lightweight 1D convolutions:
\[
\mathbf{y} = \mathbf{z} + F(\mathbf{z}),
\]
where \(F(z)\) consists of two 1D convolutional layers with reduced intermediate channel width, each followed by batch normalisation and activation. This design reduces the number of parameters while preserving the benefits of residual learning.  
The block ends with max-pooling (factor~2) and dropout with rate~0.3.

The \textbf{third block} increases the number of channels from 32 to 64 using a bottleneck-style expansion module composed of three 1D convolutions:
\begin{equation}
\mathbf{y} = \mathrm{BN}_3\!\left(W_3 * \sigma\!\left(\mathrm{BN}_2\!\left(W_2 * \sigma\!\left(\mathrm{BN}_1\!\left(W_1 * \mathbf{x}\right)\right)\right)\right)\right),
\end{equation}
where \(W_1\) and \(W_3\) are 1D convolutions with kernel size~1, and \(W_2\) is a 1D convolution with kernel size~3. Let \(\mathrm{BN}_i\) denote the batch normalisation applied after the \(i^{\text{th}}\) convolution. 
The first two convolutions are each followed by batch normalisation and ReLU activation; the final convolution is followed only by batch normalisation.  
This design enables parameter-efficient widening by reducing the number of channels in the intermediate layers, avoiding the cost of a wide convolution.  
The block concludes with max-pooling (factor~2) and dropout with rate~0.3.

The \textbf{classification head} applies global average pooling across the temporal dimension, reducing the output tensor to a 64-dimensional feature vector by averaging across the temporal axis.
A fully connected layer (64\(\to\)1) produces the binary classification logit without an activation function, as the model is trained using a loss function that incorporates the sigmoid non-linearity.

This description provides all hyperparameters, activation placements, and custom-module definitions required to reproduce the model exactly.

\subsection{Training and Optimisation}

The previously described architecture was trained to classify traces as either pure noise (background) or containing simulated cosmic-ray pulses (signal), as defined in Section~\ref{sec:dataset_definition}.
The loss function was the binary cross-entropy with logits,
\begin{equation}
\mathcal{L}_{\mathrm{BCE}} = - \frac{1}{N} \sum_{i=1}^N \left[ y_i \log \sigma(z_i) + (1 - y_i) \log\left( 1 - \sigma(z_i) \right) \right],
\end{equation}
where \( z_i \) is the model’s raw output (logit), \( y_i \) the binary target label, \( \sigma(z) \) the sigmoid function, and \( N \) the mini-batch size.

The model was optimised using the Adam algorithm \cite{kingma2014adam} as implemented in PyTorch, 
with a fixed learning rate of \( \eta = 10^{-3} \), and default values for the exponential decay rates and numerical stability parameter: 
\( \beta_1 = 0.9 \), \( \beta_2 = 0.999 \), and \( \epsilon = 10^{-8} \).
Training was performed for 15\,epochs, and the model state corresponding to the lowest validation loss was retained for evaluation.

The dataset used for model training and evaluation consisted of normalised, fixed-length radio traces drawn from two equally sized classes, corresponding to the \textbf{training and cross-validation datasets} defined in Section~\ref{sec:dataset_definition}.
The dataset was randomly shuffled and split into training and validation subsets. 
Training mini-batches were drawn from both classes in equal proportion to avoid class imbalance.

To improve robustness for hardware-friendly deployment, unstructured magnitude pruning was applied to all convolutional layers.  
A polynomial schedule gradually increased weight sparsity from zero to 50\% between early and mid-training epochs:
\begin{equation}
s(e) =
\begin{cases}
0, & e < e_{\mathrm{start}},\\[3pt]
s_{\mathrm{final}} \cdot \left( \frac{e - e_{\mathrm{start}}}{e_{\mathrm{end}} - e_{\mathrm{start}}} \right)^2, & e_{\mathrm{start}} \le e \le e_{\mathrm{end}},\\[3pt]
s_{\mathrm{final}}, & e > e_{\mathrm{end}},
\end{cases}
\end{equation}
where \( s(e) \) denotes the sparsity at epoch \( e \), \( s_{\mathrm{final}} = 0.5 \) is the target sparsity, and \( e_{\mathrm{start}} \) and \( e_{\mathrm{end}} \) are the pruning schedule boundaries.  
During training, pruning masks were dynamically updated, and after training completed, masks were removed to yield the final pruned model.

After each epoch, the network was evaluated on the validation set without gradient updates. 
The primary selection criterion for saving the best model was the validation loss. 
For monitoring purposes during development, the area under the receiver operating characteristic (ROC) curve (AUC) was also computed using a fixed classification threshold of 0.5.

The complete training pipeline, including dataset preparation, augmentation through pulse injection, pruning schedule, and validation, was implemented in PyTorch and executed on CUDA-enabled GPUs.

\section{Model Performance Analysis (Float Model)}
\label{sec:modelperformance}

To evaluate the classifier's final performance in a realistic triggering context, I consider 
\textbf{signal traces from Validation Dataset~1} and 
\textbf{background traces from Validation Dataset~3}.
Validation Dataset~1 contains noise traces with injected MGMR3D-simulated air-shower pulses at controlled SNRs matching the training distribution.
Validation Dataset~3 consists of one million measured background traces recorded in a high-interference environment.
The large sample size enables the evaluation of the model’s false-positive rejection efficiency in a highly challenging regime, where even a small false-positive rate would correspond to rates approaching the MHz range in a real-time self-trigger application.

\begin{figure}[htbp]
\centering
\includegraphics[width=0.9\textwidth]{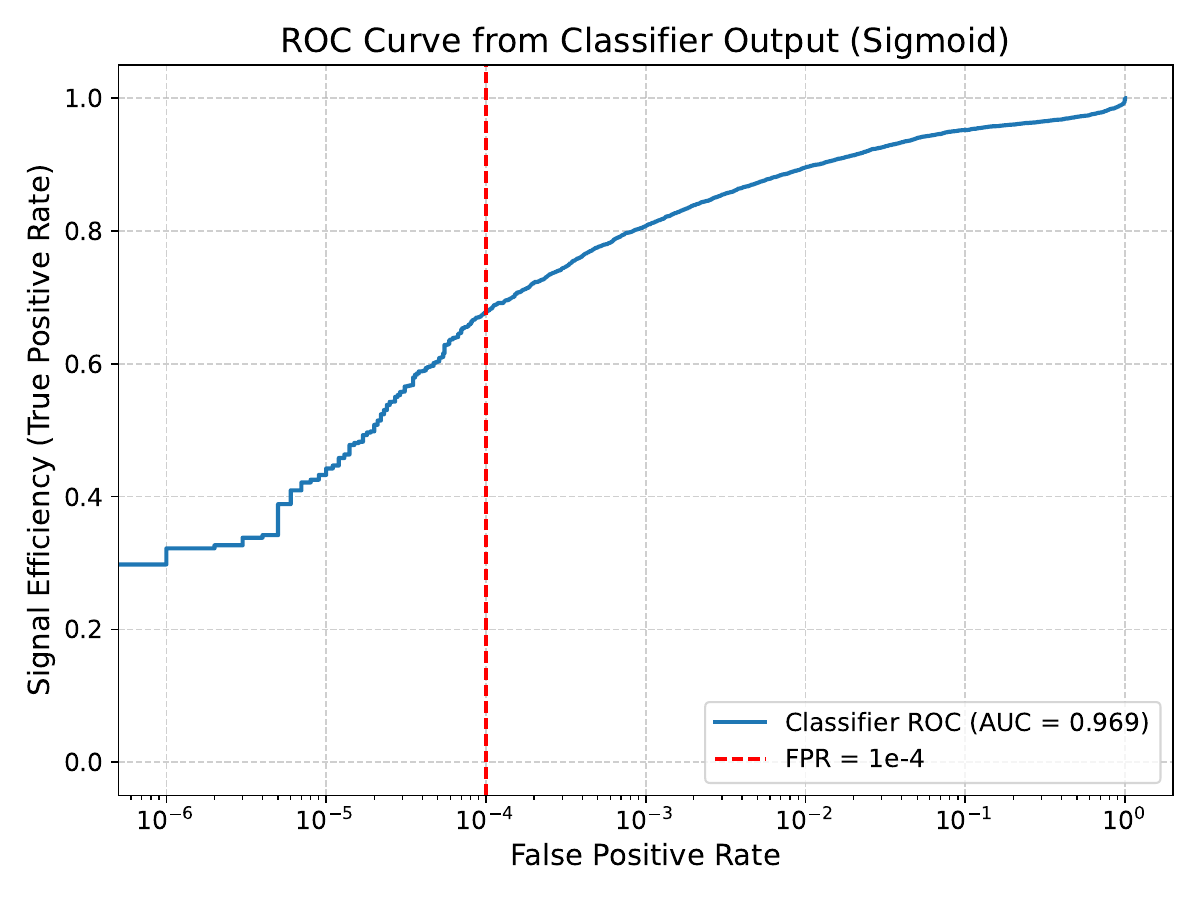}
\caption{ROC curve for the classifier evaluated on signal from Validation 1 and background from Validation 3. The solid line shows the model response; the area under the curve (AUC) is 0.969. The vertical red dashed line indicates the chosen operating point at a false positive rate of $10^{-4}$.}
\label{fig:roc_hist}
\end{figure}

The main performance indicator is the \textit{receiver operating characteristic} (ROC) curve (Figure~\ref{fig:roc_hist}), obtained from classifier outputs for signal traces from Validation~1 and background traces from Validation~3. 
The solid curve shows the trade-off between the true positive rate (signal efficiency) and the false positive rate as the classification threshold is varied, with an area under the curve (AUC) of~0.969. 
The vertical red dashed line marks the chosen operating point at a false positive rate of \(10^{-4}\), illustrating that the model maintains high efficiency while suppressing false triggers to extremely low rates.



While the ROC analysis focuses on the nominal SNR distribution, model performance was also evaluated on \textbf{Validation Dataset~2}, where most injected pulses have SNRs well below the noise floor. At a fixed false positive rate of \(10^{-4}\), the signal efficiency decreases from \(68\%\) on Validation~1 to \(27\%\) on Validation~2. This drop reflects the expected trend toward lower classification scores for weaker signals. These results provide a first indication that the network retains some discriminative power in the low-SNR regime.

\subsection{False Trigger Rate Estimation in a High-Noise Environment}

To assess the model’s robustness in realistic high‑noise conditions, I evaluated it on \textbf{Validation Dataset~3}.  
This dataset reflects high‑RFI environments such as those encountered in urban settings (\textit{e.g.}, LOFAR, LOPES) or at certain AERA stations~\cite{Schmidt2011-xy}.  
Such conditions are representative of operational scenarios in which a self‑triggering station must continuously process high‑rate background waveforms, potentially at sampling rates of up to \(1\ \mathrm{MHz}\) trigger trials. In this regime, even a small false positive rate (FPR) translates into a substantial trigger load: for example, an \(\mathrm{FPR} = 10^{-4}\) corresponds to \(\sim 100\ \mathrm{Hz}\) first‑level trigger (FLT) rate per station.

From the ROC analysis (Figure~\ref{fig:roc_hist}), the model achieves a signal efficiency of \(68\%\) at this target operating point.
The \SI{100}{\Hz} FLT rate is already suitable for a station‑level trigger in a cosmic‑ray radio array and provides a strong foundation for further refinement.
In practice, a subsequent second‑level trigger (SLT) stage—based on array‑level coincidence requirements—can further suppress false positives while preserving nearly all genuine events.  
Such multi-stage triggering strategies are widely used in cosmic-ray experiments (e.g., Auger, AERA, GRAND) and are essential for handling high raw data rates in a scalable and efficient manner.

For applications where further reduction of the FLT rate is desired, the threshold can be tightened to \(\mathrm{FPR} = 10^{-6}\), corresponding to roughly \SI{1}{\Hz} at the station level.  
This operating point still retains a signal efficiency of \(32\%\) and may be useful in scenarios where exceptionally low background rates are advantageous.  
In practice, \(\mathrm{FPR} = 10^{-4}\) already offers a good balance between efficiency and manageable trigger rates, with the lower‑FPR setting available as an option when background suppression takes priority.

\subsection{Example of a Low-Amplitude Signal Classification}

\begin{figure}[htbp]
\centering
\includegraphics[width=0.9\textwidth]{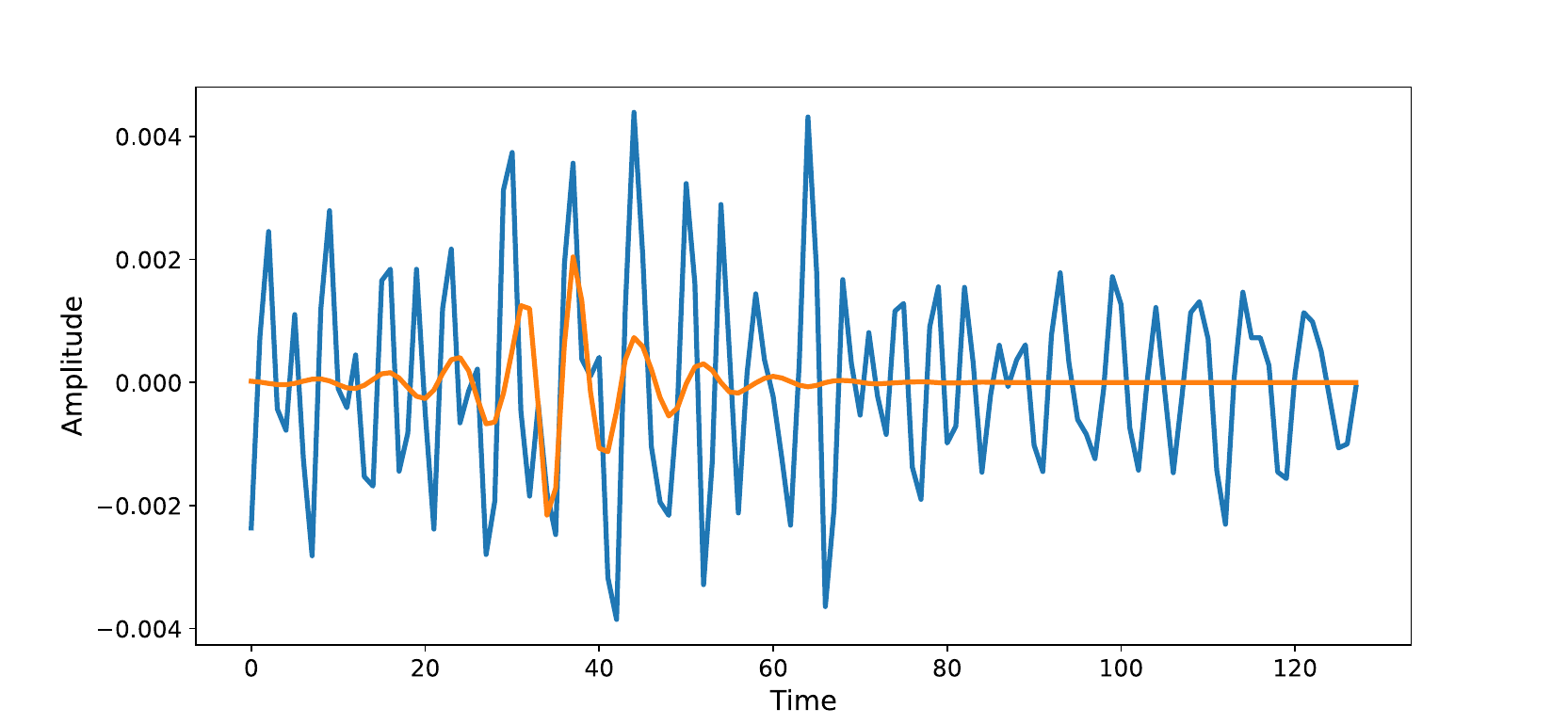}
\caption{Representative trace with a faint signal pulse correctly classified by the model despite transient noise, demonstrating its ability to recover weak air shower signals undetectable by threshold-based methods.}
\label{fig:faint_pulse}
\end{figure}

Figure~\ref{fig:faint_pulse} illustrates a \textbf{representative trace containing a faint signal pulse} correctly classified by the model. The pulse is indistinguishable by eye due to transient noise, yet the classifier confidently assigns it a high probability. This result highlights the model’s ability to recover weak air shower signals that would otherwise remain undetectable using traditional threshold-based methods.

\subsection{Comparison with a Threshold-Based Trigger}
\label{sec:baseline_threshold}

To provide a baseline against conventional self-triggering techniques, I implemented a simple \emph{SNR-threshold trigger}. 
For each evaluation trace, I form the analytic signal via the Hilbert transform and define the score as in Eq.~\ref{eq:snr}.
For background traces, the trace itself serves as this reference, so the largest envelope excursion is treated as a spurious candidate signal. 
The resulting distributions of scores for signal and background events allow the construction of a ROC curve by varying the threshold on the SNR score.

To ensure a fair comparison, both the neural-network classifier and the SNR-threshold baseline were evaluated on the same unfiltered datasets, which contain the full interference environment as described in Section~\ref{sec:dataset_definition}. 
In practice, threshold-based self-triggers are often preceded by frequency filtering to increase sensitivity, but such preprocessing would also be expected to benefit the neural network if applied consistently. 
Applying additional filtering only to the baseline would artificially simplify its input and alter the effective noise conditions relative to the classifier evaluation. For this reason, I restrict the comparison to identical unfiltered inputs for both methods.

For a fixed input, ROC performance is invariant under any monotonic reparameterisation of the score. 
For example, using the peak-to-RMS ratio or the z-score is mathematically equivalent to our SNR definition up to a monotonic transform and therefore yields the same ROC curve within numerical precision. By contrast, using the raw peak voltage without noise normalisation performs significantly worse, as expected, because it does not account for fluctuations in the noise level. More elaborate variants (e.g.\ running-average baselines, short-time energy) remain threshold triggers in spirit but introduce method-specific hyper-parameters and design choices. To avoid cherry-picking, I therefore restrict the comparison to this minimal representative baseline.

\begin{figure}[htbp]
  \centering
  \includegraphics[width=0.9\textwidth]{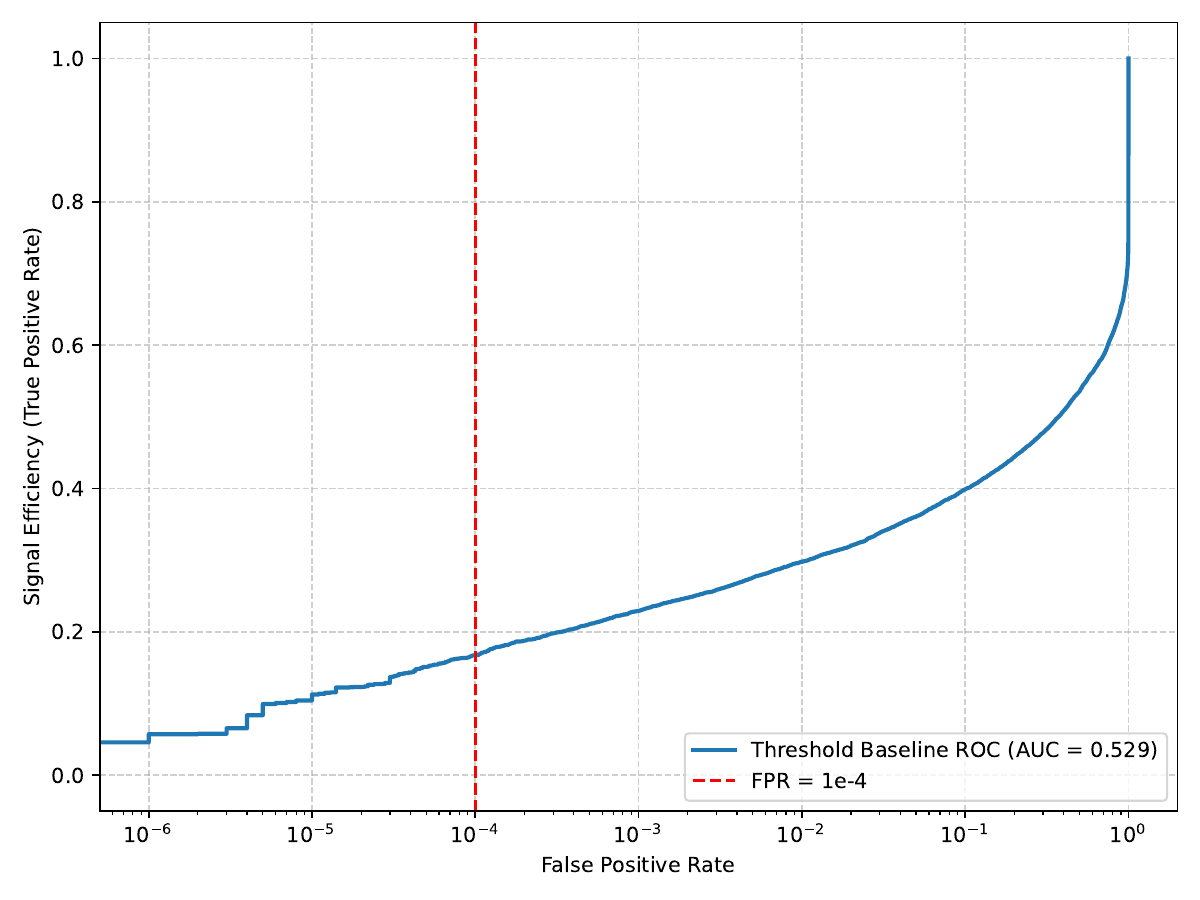}
  \caption{ROC curve for the SNR-threshold baseline (Hilbert-envelope peak squared divided by noise variance). The area under the curve is \(0.529\). The vertical red dashed line indicates the operating point at a false positive rate of $10^{-4}$; the corresponding signal efficiency is \(\approx 16.8\%\).}
  \label{fig:roc_baseline}
\end{figure}

Figure~\ref{fig:roc_baseline} shows the ROC of the SNR-threshold trigger. 
The area under the curve is \(\mathrm{AUC}=0.529\). 
At the operating point used throughout this work, corresponding to a false positive rate of \(10^{-4}\) (vertical red dashed line), the threshold-based trigger achieves a signal efficiency of only \(\approx 16.8\%\). 
For comparison, the neural-network classifier maintains an efficiency of \(\sim 68\%\) under the same conditions (cf.\ Fig.~\ref{fig:roc_hist}). 
This demonstrates a clear improvement of the AI-based approach over a traditional threshold trigger in the challenging high-interference environment considered here.

\subsection{Physics Implications of Improved Triggering}
\label{sec:physics_impact}

Beyond the technical demonstration, it is instructive to consider the physics impact of lowering the trigger threshold in realistic, interference-rich environments. 
At the chosen operating point (false positive rate of $10^{-4}$), the SNR-threshold baseline achieves a signal efficiency of only $\sim 16.8\%$ (Sec.~\ref{sec:baseline_threshold}), whereas the neural-network classifier retains an efficiency of $\sim 68\%$ (Sec.~\ref{sec:modelperformance}). 
This corresponds to an effective reduction of the detectable signal-to-noise threshold by more than a factor of two.

Such an improvement has direct consequences for the reach of radio arrays. Lowering the trigger threshold extends the accessible energy range towards lower primary energies, where air-shower radio signals are intrinsically weaker. It also improves sensitivity to very inclined air showers, since the particle component is largely absorbed in the atmosphere while the radio emission remains strong. These events are particularly valuable for composition studies, for enlarging the sky coverage of ground-based observatories, and as a prerequisite for detecting ultra-high-energy neutrinos, which manifest as upward-going or near-horizontal showers. By enabling robust radio-only operation under high-background conditions, the method reduces reliance on external particle detectors, thereby simplifying array design and allowing sparser station layouts. This opens a path toward larger effective apertures at EeV energies, complementing and extending the capabilities of existing cosmic-ray and neutrino observatories.

These considerations show that the proposed AI-based trigger is not only technically feasible on FPGA hardware, but also provides a tangible physics benefit by increasing the accessible parameter space of radio arrays in energy, inclination, and coverage.

\section{Quantisation and FPGA Implementation}

\subsection{Quantisation with HLS4ML}
For real-time deployment, the trained floating-point classifier was converted to a fixed-point representation using the \texttt{HLS4ML} framework~\cite{fastml_hls4ml,Duarte:2018ite,Aarrestad:2021zos}.  
Quantisation replaces floating-point multiplications and additions with fixed-point arithmetic, reducing resource usage and latency on FPGAs while maintaining compatibility with hardware synthesis workflows.  
The key challenge is to choose a precision that minimises resource cost without sacrificing classification performance.

To determine a suitable fixed-point format for FPGA deployment, I first estimated the minimum required integer bit-width to avoid overflow. This was done by forwarding representative validation samples through the trained model while recording the maximum absolute activation values across all intermediate layers. Based on this analysis, I fixed the integer width at 5 bits.

Subsequently, I scanned the total bit-width while keeping the integer part fixed, evaluating classification performance at each precision setting. Figure~\ref{fig:accuracy_fpga}a shows the area under the ROC curve (AUC) as a function of total bit-width. Performance remains stable down to 13 total bits, below which degradation becomes apparent. I therefore selected \verb|ap_fixed<13,5>| (13 total bits, 5 integer bits) as the default precision for FPGA synthesis.

\subsection{Float vs.\ Quantised Model Performance}
\begin{figure}[ht]
  \centering
  \begin{subfigure}[t]{0.495\textwidth}
    \centering
    \includegraphics[height=4.5cm]{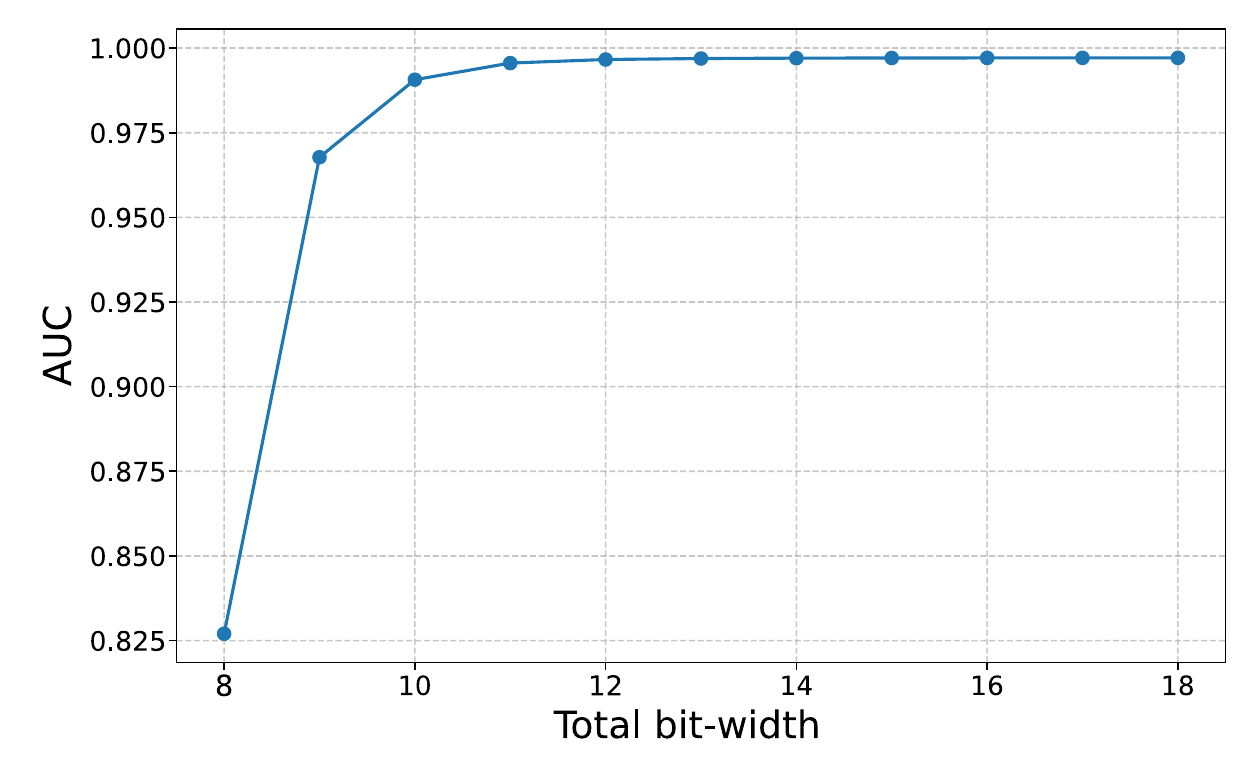}
    \caption{AUC versus total bit-width (fixed integer width: 5 bits).}
    \label{fig:bitscan}
  \end{subfigure}
  \hfill
  \begin{subfigure}[t]{0.495\textwidth}
    \centering
    \includegraphics[height=4.5cm]{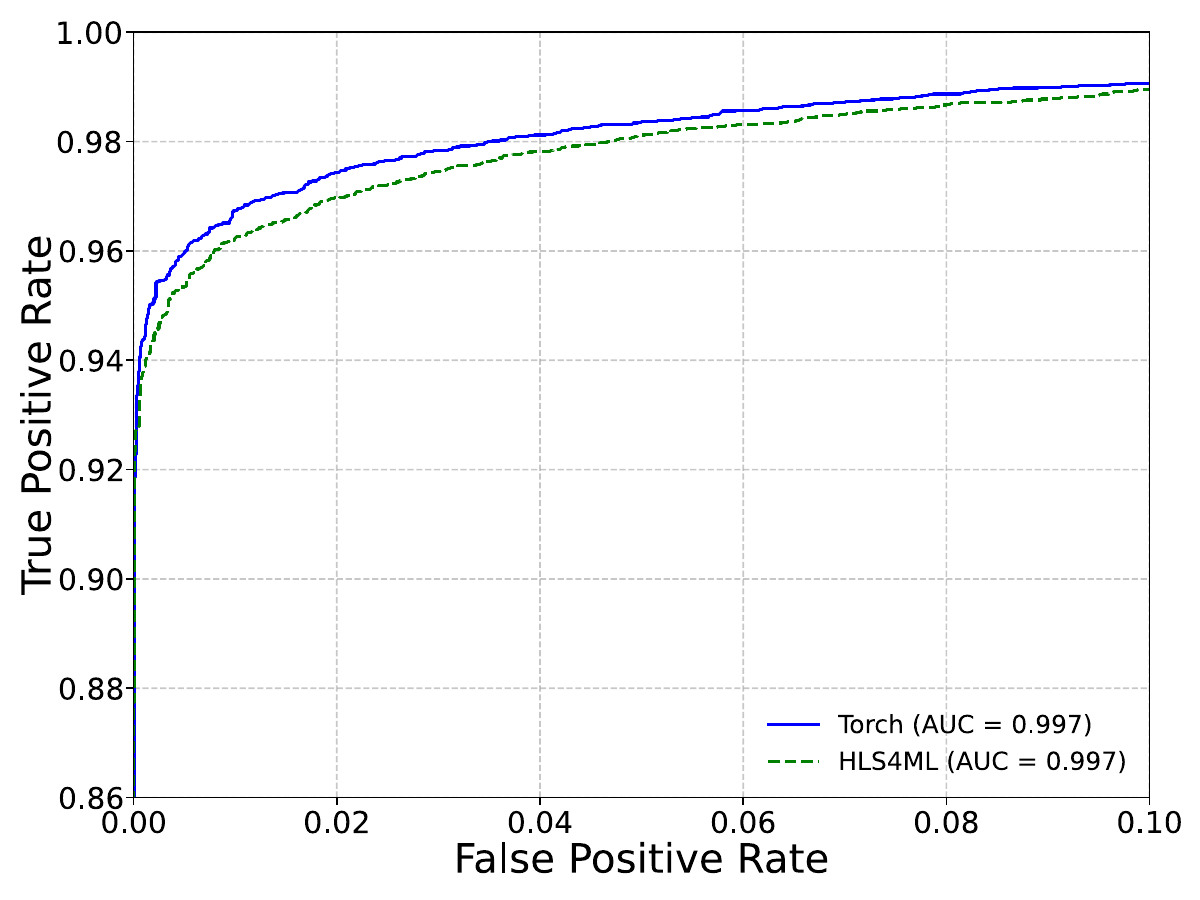}
    \caption{ROC comparison of the floating-point and quantised classifiers.}
    \label{fig:roc}
  \end{subfigure}
  \caption{Accuracy characteristics of the fixed-point implementation.}
  \label{fig:accuracy_fpga}
\end{figure}

The effect of quantisation on classification performance was evaluated by comparing the floating-point and \verb|ap_fixed<13,5>| implementations on the same validation datasets.  
As shown in Figure~\ref{fig:accuracy_fpga}b, the ROC curves for the two models are nearly identical, with the AUC decreasing only marginally from the PyTorch floating-point baseline (\(\mathrm{AUC}_{\mathrm{float}} = 0.997\)) to the quantised implementation (\(\mathrm{AUC}_{\mathrm{quant}} = 0.996\)).  
This confirms that the chosen fixed-point precision preserves the model’s separation power between signal and background, enabling an efficient FPGA implementation without loss of physics performance.

\subsection{FPGA Synthesis with Vitis~2023}
The quantised model was synthesised using \texttt{Vitis HLS 2023.2}~\cite{vitisHLS2023}, targeting three representative FPGA families: the low-cost Zynq-7000 (xc7z020), the mid-range Kintex UltraScale (xcku040), and the high-end Alveo U250 (xcu250).  
A target clock period of \SI{5}{\nano\second} (\SI{200}{\mega\hertz}) was assumed in all cases.  
The measured end-to-end latency, pipeline initiation interval (II), and DSP usage are summarised in Table~\ref{tab:fpga_latency_real}.  

\begin{table}[ht]
  \centering
  \caption{Measured latency and DSP usage for the \texttt{ap\_fixed<13,5>} implementation. DSP usage is reported as a fraction of available slices; it is the primary limiting resource in this design.}
  \label{tab:fpga_latency_real}
  \begin{tabular}{lcccc}
    \toprule
    Platform & Latency [$\mu$s] & II & DSP usage & Timing met \\
    \midrule
    Zynq-7000 xc7z020          & 6.51 & 1060 & 963\,\% & yes  \\
    Kintex UltraScale xcku040  & 3.30 &  654 & 108\,\% & yes \\
    Alveo U250 xcu250          & 1.99 &  394 &  17\,\% & yes \\
    \bottomrule
  \end{tabular}
\end{table}

\subsubsection{Resource Utilisation}
DSP slices are the critical limiting resource for this design, as they are heavily used for multiply–accumulate (MAC) operations in convolutional and dense layers.  
While LUT and BRAM usage remain well below device limits in all cases, DSP saturation is encountered for the Zynq-7000 and, to a lesser extent, the Kintex UltraScale.  
This constraint dictates optimisation strategies, such as resource sharing or further model compression, for deployment on cost-sensitive platforms.  
The Alveo U250, by contrast, offers substantial headroom for scaling up the model or reducing latency further.

\subsubsection{Latency}
\begin{figure}[ht]
  \centering
  \includegraphics[width=.95\textwidth]{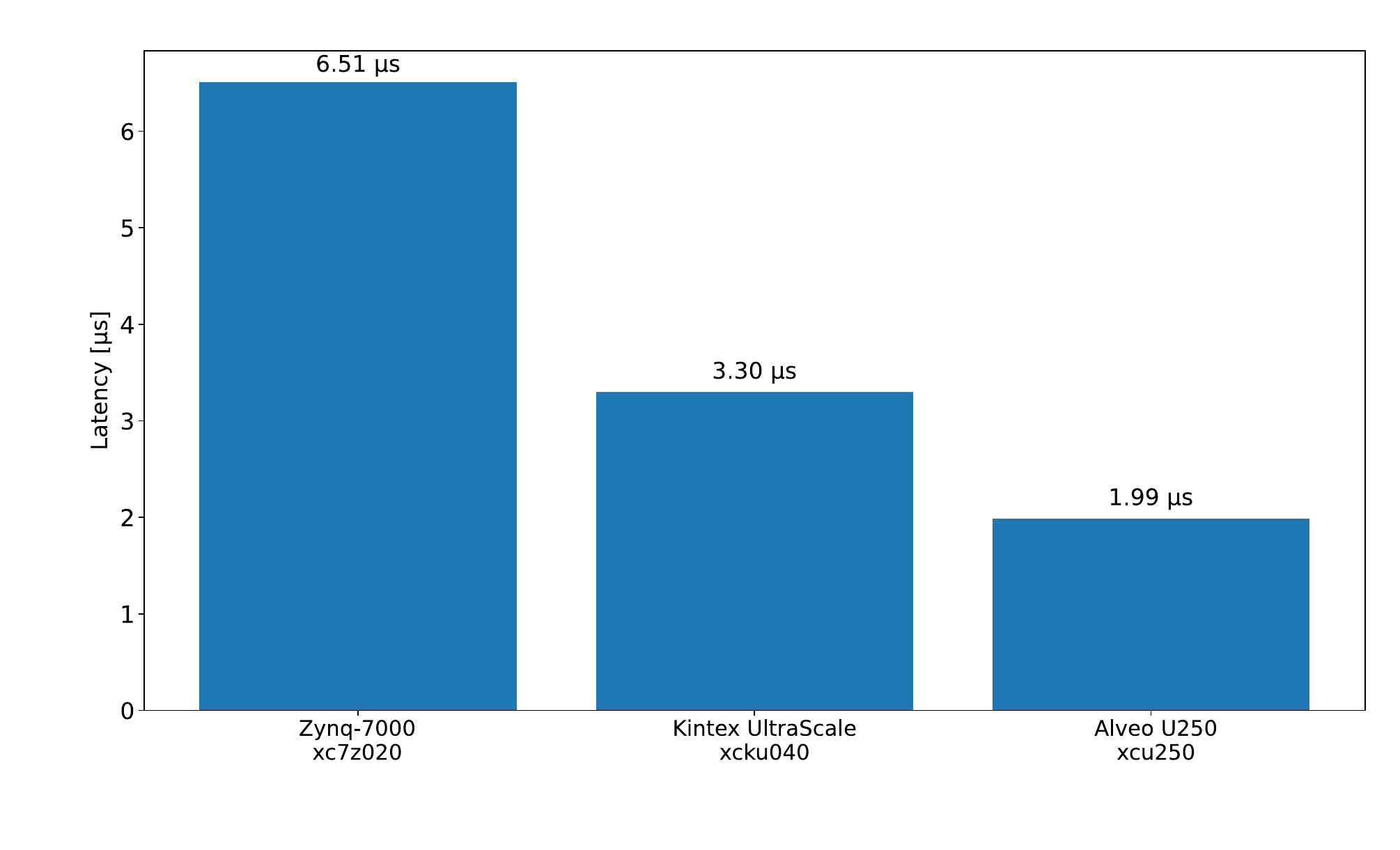}
  \caption{Measured inference latency for the three FPGA targets. All meet the \(\mathcal{O}(\mu\mathrm{s})\) requirement for first-level triggering.}
  \label{fig:latency_fpga}
\end{figure}

Figure~\ref{fig:latency_fpga} shows the measured inference latency for the three synthesis targets.  
Even without aggressive optimisation, the high-end Alveo U250 achieves sub-\SI{2}{\micro\second} latency.  
The Kintex UltraScale operates at \SI{3.3}{\micro\second}, while the Zynq-7000 delivers \SI{6.5}{\micro\second}.  
All results meet the \(\mathcal{O}(\mu\mathrm{s})\)-level latency guideline for station-level triggering in the target application.

\subsection{Power consumption}

I characterised \emph{fabric-only} on-chip power for the \texttt{HLS4ML}-generated accelerator implemented on a \texttt{Xilinx Alveo U250 (xcu250-figd2104-2L-e)} using \texttt{Vivado~2023.2}. To measure only the neural-network accelerator’s contribution, I built a minimal ‘fabric-only’ design: the \texttt{HLS4ML} core was the sole processing block at the top level, with no platform shell, memory controllers, PCIe or other peripherals, and no external~I/O interfaces. The only additional logic was an on-chip pseudo-random bit sequence (PRBS) generator to provide a continuous input stimulus. The design was driven by a single \SI{200}{\mega\hertz} clock and taken through full implementation (synthesis, placement, and routing) so power was reported on the placed-and-routed netlist.

The PRBS source was implemented as a 32-bit LFSR with taps $x^{32}{+}x^{22}{+}x^{2}{+}x{+}1$, replicated across the input bus, with handshake/control pins held active to keep the pipeline filled. This stimulus exercises the full datapath and prevents constant-propagation through the network.

Post-implementation power was obtained with Vivado’s vectorless analysis (no activity annotation). The placed-and-routed netlist reports a dynamic power of \textbf{\SI{5.51}{\watt}} for the core. If clock-network power is also excluded, the dynamic component attributable solely to datapath logic is approximately \SI{5.06}{\watt}. Because the estimate remains vectorless and excludes platform infrastructure, the reported numbers should be interpreted as a \emph{lower bound}; absolute power is expected to increase with activity-annotated runs and board-level I/O. As the U250 is the largest of the three target platforms, this value is conservative with respect to device scaling.

These results indicate ample headroom within a typical $\sim\SI{15}{\watt}$ budget for a compact radio-detection pipeline, leaving margin for front-end filtering, data conversion, and control logic. A full platform-inclusive power profile (including activity-annotated power and comparisons of LUT- vs.\ DSP-mapped arithmetic) is left for future hardware studies.

\section{Summary and Outlook}
\label{sec:conclusion}

I have presented an artificial‑intelligence‑based self‑trigger for radio detection of extensive air showers and demonstrated its feasibility for real‑time deployment on FPGA hardware.  
Starting from a floating‑point model trained in \texttt{PyTorch} on measured noise from a high‑interference environment combined with cosmic‑ray pulses generated using the MGMR3D simulation framework, the workflow covers a broad range of shower geometries, energies, and antenna--core distances.  
The trained fully convolutional classifier achieves stable performance under operating conditions representative of MHz‑scale trigger‑trial rates encountered in urban or RFI‑rich sites.

For context, I also compared the neural-network classifier to a simple threshold-based trigger, 
implemented as an SNR cut on the Hilbert-envelope peak relative to the noise variance. 
The baseline achieves only modest separation power (\(\mathrm{AUC}=0.529\)) and a signal efficiency of 
\(\sim 17\%\) at the target operating point of \(\mathrm{FPR}=10^{-4}\), 
whereas the classifier retains \(\sim 68\%\) efficiency under the same conditions. 
This comparison demonstrates that the AI-based approach provides a substantial improvement over 
traditional threshold triggers in the challenging high-interference environment considered here.

The network was quantised using \texttt{HLS4ML} to a fixed‑point precision of \verb|ap_fixed<13,5>|.  
A precision scan confirmed that 13‑bit fixed‑point arithmetic preserves the physics performance of the floating‑point model (\(\mathrm{AUC}_{\mathrm{float}} = 0.997\), \(\mathrm{AUC}_{\mathrm{quant}} = 0.996\)).  
The quantised model was synthesised with \texttt{Vitis HLS~2023.2} for three FPGA platforms (Zynq‑7000, Kintex UltraScale, and Alveo U250).  
High‑level synthesis results show inference latencies in the microsecond range, meeting the first‑level trigger timing budget for large‑scale radio arrays.  
DSP usage emerged as the primary resource bottleneck, guiding optimisation strategies for deployment on cost‑constrained devices. A post-implementation, fabric-only power estimate on the Alveo U250 reports \SI{5.51}{\watt} at \SI{200}{\mega\hertz} (vectorless analysis), which I interpret as a conservative lower bound on on-chip power.

For the datasets discussed in this work, strong narrow‑band RFI components were not removed prior to training.  
In a real‑world deployment, eliminating such RFI using an FFT‑based filter implemented on the FPGA could simplify the classification task by reducing background complexity.  
A simpler network architecture may lead to faster inference times and lower computational costs, paving the way for real‑time triggering with further reduced latency and resource usage.

This work demonstrates, for the first time in the context of radio-detector self-triggering, an end-to-end workflow from AI model training to FPGA-oriented deployment using \texttt{HLS4ML}: quantisation to fixed-point, high-level synthesis with \texttt{Vitis HLS~2023.2}, and validation of physics-performance preservation together with post-synthesis latency/resource characterisation and a fabric-only, post-implementation power estimate. 
The approach reduces reliance on external triggers from particle or fluorescence detectors and supports autonomous operation scenarios for next-generation radio arrays.

Future developments  include broadening the training to a wider family of simulated pulse morphologies and incorporating additional measured air-shower events. On the hardware side, it will be important to complement the fabric-only power estimate reported here with activity-annotated and platform-inclusive profiles across the target devices, and to investigate architectural and clocking optimisations suitable for low-power autonomous stations. Together with simple on-FPGA pre-filtering of narrow-band RFI, these directions are expected to further reduce resource usage and latency and to de-risk deployment in operational radio arrays.

In addition to these technical aspects, the comparison with a simple SNR-threshold baseline highlights the physics impact of the proposed method. At the chosen operating point, the neural-network trigger lowers the effective trigger threshold significantly compared to the SNR baseline. This improved threshold increases sensitivity to weak signals and to very inclined air showers, where the particle cascade is absorbed while the radio emission remains detectable. Such events are essential not only for composition studies and for extending the sky coverage of cosmic-ray observatories, but also as a prerequisite for detecting ultra-high-energy neutrinos with ground-based radio arrays. Moreover, the approach is designed to operate robustly in environments with strong and variable radio backgrounds, where conventional threshold triggers fail, thereby enabling autonomous operation under realistic field conditions. These results demonstrate that FPGA-based AI self-triggering not only meets the technical feasibility requirements but also provides a tangible enhancement of the scientific potential of next-generation radio and neutrino observatories.

\acknowledgments
The author gratefully acknowledges the support of the \textbf{Electronics Laboratory (E-Lab)} of the Physics Department at Universität Siegen for their assistance in designing and producing the electronics used in this research. Their contributions facilitated the development and testing of the hardware components, streamlining the experimental setup.

The author also gratefully acknowledges Prof. Katie Mulrey and Prof. Harm Schoorlemmer (Radboud University Nijmegen) for kindly providing the antenna employed in this study. The assistance of Noah Siegemund in setting up the hardware on campus is also gratefully acknowledged.

This research made use of open-source software, including \textbf{PyTorch}, \textbf{HLS4ML} and various scientific computing libraries in Python, such as \textbf{NumPy, SciPy, and Matplotlib}. The author acknowledges their developers for providing these valuable tools, which played a crucial role in data processing, model training, and analysis.


\bibliographystyle{JHEP}  
\bibliography{paper}







\end{document}